\documentclass[runningheads]{llncs}

\usepackage{url}
\usepackage{color}
\usepackage{tabularx}
\usepackage{cite}

\newcommand{\q}[1]{\lq\lq{}{}#1\rq\rq{}{}}

\begin{document}

\title{Time-Aware and Corpus-Specific\\Entity Relatedness}

\author{
	Nilamadhaba Mohapatra\inst{1,2} \and 
    Vasileios Iosifidis\inst{1} \and 
    Asif Ekbal\inst{2} \and \\
    Stefan Dietze\inst{1} \and 
    Pavlos Fafalios\inst{1} }
\institute{
    L3S Research Center, University of Hannover, Germany\\
    \email{\{iosifidis, dietze, fafalios\}@L3S.de}
    \and
   Indian Institute of Technology, Patna, India\\
    \email{\{nilamadhaba.mtmc15, asif\}@iitp.ac.in}
}

\authorrunning{N. Mohapatra et al.}

\maketitle

\begin{abstract}
Entity relatedness has emerged as an important feature in a plethora of applications such as information retrieval, entity recommendation and entity linking.
Given an entity, for instance a person or an organization, entity relatedness measures can be exploited for generating a list of highly-related entities. However, the relation of an entity to some other entity depends on several factors, with \emph{time} and {\em context} being two of the most important ones (where, in our case, {\em context} is determined by a particular corpus). For example, the entities related to the \emph{International Monetary Fund} are different now compared to some years ago, while these entities also may highly differ in the context of a USA news portal compared to a Greek news portal. 
In this paper, we propose a simple but flexible model for entity relatedness which considers time and entity aware word embeddings by exploiting the underlying corpus. The proposed model does not require external knowledge and is language independent, which makes it widely useful in a variety of applications.  


\end{abstract}

\section{Introduction}
\label{sec:intro}

Entity relatedness is the task of determining the degree of relatedness between two entities. Measures for entity relatedness facilitate a wide variety of applications such as information retrieval, entity recommendation and entity linking. Traditional approaches consider the structural similarity in a given graph and lexical features \cite{strube2006wikirelate,Milne08aneffective,hoffart2012kore,nunes2013combining,hulpucs2015path,ponza2017two}, or make use of Wikipedia-based entity distributions and embeddings \cite{aggarwal2014wikipedia,basile2016learning}.

Recent works on related topics have supported the hypothesis that the context of entities is temporal in nature. Specifically, \cite{FangC14} has shown that prior probabilities often change across time, while \cite{zhang2016probabilistic} showed that the effectiveness of entity recommendations for keyword queries is affected by the time dimension.
In addition, \cite{tran2017beyond} introduced the notion of contextual entity relatedness and showed that entity relatedness is both time and context dependent. In that work, context refers to {\em topic} ({\em aspect}), i.e., the goal is to find the most related entities given an entity and an aspect (e.g., {\em relationship} or {\em career} are two different aspects for a person).
Finally, \cite{prangnawarattemporal} showed that exploiting different temporal versions of a knowledge base, each reflecting a different time period, can improve the accuracy of entity relatedness.

On the other hand, entity relatedness is also strongly dependent on the {\em corpus context} at hand. For instance, given a search application operating over a collection of German news articles from  summer 2014 and the query entity {\em 2014 FIFA World Cup} (\url{https://en.wikipedia.org/wiki/2014_FIFA_World_Cup}), a list of related entities should include {\em Germany national football team}, {\em Argentina national football team}, {\em Mario G\"{o}tze}, and {\em Brazil}. However, considering a collection of Greek articles of the same time period, the top related entities might include {\em Greece national football team}, {\em Costa Rica national football team}, and {\em Sokratis Papastathopoulos} (entities that are not important or popular in German news).

To this end, our work considers entity relatedness as a measure which is strongly dependent on both the time aspect and the corpus of the underlying application. Contrary to existing approaches that train embeddings on general-purpose corpora, in particular Wikipedia \cite{aggarwal2014wikipedia,basile2016learning}, we compute entity relatedness given a specific collection of documents that spans a particular time period. Our approach exploits entities extracted from the underlying corpus for building time and entity aware word embeddings. The evaluation results show that the proposed model outperforms similar time and entity agnostic models.

The remainder of this paper is organized as follows: Section \ref{sec:problemDef} defines the problem of \textit{time-aware and corpus specific entity relatedness}, Section \ref{sec:approach} details the proposed method, Section \ref{sec:eval} presents evaluation results, and finally Section \ref{sec:conclusion} concludes the paper and discusses interesting directions for future research.

\section{Problem Definition}
\label{sec:problemDef}

Let $D$ be a corpus of documents, e.g., a collection of news articles, covering the time period $T_D = [\tau_{s}, \tau_{e}]$ (where $\tau_{s},\tau_{e}$ are two different time points with $\tau_{s} < \tau_{e}$).
For a document $d \in D$, let $E_d$ denote all entities mentioned in $d$ extracted using an entity linking method \cite{shen2015entity}, where each entity is associated with a unique URI in a knowledge base like Wikipedia. This list of extracted entities may include persons, locations, etc., but also events (e.g., {\em US 2016 presidential election}) and more abstract concepts such as {\em democracy} or {\em abortion}. 
Finally, let $E_D$ denote all entities mentioned in documents of $D$.

Consider now a collection of documents $D$ covering a time period $T_D$ and a set of entities $E_D$ extracted from and prevalent in $D$. 
Given i) one or more {\em query entities} $E_q$
and ii) a time period of interest $T_q \subseteq T_D$,
the task {\em \q{time-aware and corpus-specific entity relatedness}} focuses on finding a top-$k$ list of entities $E_k \subset E_D$ related to the query entities $E_q$ in terms of both $T_q$ and $D$.
We model this task as a ranking problem, where we first generate a list of {\em candidate entities} $E_c \subset E_D$ and then rank this list of entities based on their relevance to the query entities. 
For generating the list of candidate entities, one can follow an approach similar to \cite{zhang2016probabilistic} which exploits Wikipedia links, DBpedia and entity co-occurrences in the annotated corpus, or consider the connectedness of the query entity with other entities in Social Media \cite{fafalios2017tpdl}.

\section{Approach}
\label{sec:approach}
Word embeddings provide a distributed representation of words in the semantic space.
We generate word vectors following the distributional models proposed in \cite{mikolov2013distributed,mikolov2013efficient} and the well-known Word2Vec tool.

\subsection{Time-Aware Word Vector Similarity} 

We group all documents in $D$ into time-specific subsets $C = (C_1, \dots, C_n$) based on a fixed time granularity $\Delta$ (e.g., week, month, or year). 
We then preprocess the documents of each subset $C_i \in C$ and train a Word2Vec Continuous Bag of Words (CBOW) model. 
Each model generates a matrix \textbf{$keys \times dimension(d)$} where $keys$ represents a word from the corpus $C_i$ for which a $d$-dimensional vector exists in the trained model. 

To find the relatedness score between a query entity $e_{q} \in E_q$ and a candidate entity $e_c \in E_c$, we compute the cosine similarity of their word vectors in the corresponding trained model for the input time period $T_q$. In our experiments, for finding the words that represent a query or candidate entity, we use the last part of the entity's Wikipedia URI by first replacing the underscore character with the space character and removing any text in parentheses. Based on the underlying knowledge base, one could use here other approaches, e.g., exploit the entity's label in DBpedia. 
For multi-word entities, we compute the average vector of the constituent words, while for inputs consisting of more than one entity, we compute the average of all entity vectors.

\subsection{Considering the Entity Annotations}

There are two limitations in dealing with the embedding vectors obtained in the previous model: (i) for multi-word entities the average of the word vectors of the constituent tokens is computed, i.e., n-grams are ignored (consider for example the entity {\em United Nations}), and (ii) the same entity mention (surface form) may refer to different entities (e.g., \textit{Kobe} may refer to the basketball player \textit{Kobe Bryant} or the \textit{Japanese city}).
To cope with these problems, we exploit the entity annotations $E_D$. Recall that each extracted entity is associated with a unique URI in a knowledge base (e.g., Wikipedia or DBpedia). As also suggested in \cite{mikolov2013distributed} for the case of phrases, we preprocess the documents and replace the entity mentions with unique IDs. The modified text corpora are now used for training the Word2Vec model, where the word vector for an entity is now calculated using its unique ID.

Now, the similarity score between a query entity $e_{q} \in E_q$ and a candidate entity $e_c \in E_c$ is the cosine similarity of their word vectors using the corresponding modified (with entity IDs) Word2Vec model for the input time period $T_q$. Formally:
\begin{equation}
\label{formula1}
\small
Sim(e_{q},e_{c},T_q)=cosine(wordvec(id(e_{q}), T_q), wordvec(id(e_{c}),T_q))
\end{equation}
where $id(e)$ is the ID of entity $e$, and $wordvec(w, T_q)$ returns the word vector of token $w$ using the corresponding model for the input time period $T_q$.

\subsection{Relaxing the Time Boundaries} 
An important event related to the query entities may happened very close to the boundaries of the time period of interest $T_q$. This means that two entities might be highly related some time before or after $T_q$. To cope with this problem, the similarity score can also consider the Word2Vec models for the time periods before and after $T_q$.
Let $T_{q}^{-1}$ and  $T_{q}^{+1}$ be the time periods of granularity $\Delta$ before and after $T_{q}$, respectively.  
The similarity score between a query entity $e_{q} \in E_q$ and a candidate entity $e_c \in E_c$ can now be defined as follows:

\begin{equation}
\label{formula2}
\small
Sim'(e_{q},e_{c},T_q) = Sim(e_{q},e_{c},T_q) \cdot w_1 + Sim(e_{q},e_{c}, T_{q}^{-1}) \cdot w_2 + Sim(e_{q},e_{c},T_{q}^{+1}) \cdot w_3  
\end{equation}

where $w_1$, $w_2$ and $w_3$ are the weights of the models of $T_q$, $T_{q}^{-1}$ and $T_{q}^{+1}$, respectively, with $w_1 + w_2 + w_3 = 1.0$.  

This modification can increase the ranking of an important candidate entity that co-occurs frequently with a query entity some time before or after $T_q$, but at the same time can decrease the ranking of an entity co-occurring with a query entity during $T_q$. Thus, we should avoid using a very small $w_1$ value.

\section{Evaluation Results}
\label{sec:eval}

Our objective is to evaluate the effectiveness of the proposed approach and compare it with similar but time and entity agnostic models.
We use the dataset and ground truth provided by \cite{zhang2016probabilistic} for the problem of time-aware entity recommendation\footnote{\url{http://km.aifb.kit.edu/sites/ter/}}.
The dataset provides candidate entities and relevance judgments for 22 keyword queries, where each query corresponds to a particular date range (month). The dataset also provides a set of more than 8M news articles spanning a period of 7 months (Jul'14-Jan'15), annotated with Wikipedia entities. 
For each of the keyword queries, we manually specified the corresponding entities (since, in our problem, the input is an unambiguous entity URI). For instance, for the query {\em Tour de France Nibali} (with date range 07/2014), the query entities are: {\em Tour de France} (\url{https://en.wikipedia.org/wiki/Tour_de_France}) and {\em Vincenzo Nibali} (\url{https://en.wikipedia.org/wiki/Vincenzo_Nibali}).
We also remove the query entities from the set of candidate entities since these are the input in our entity relatedness problem. 

We build 7 CBOW models using the proposed approach for each month from Jul'14 to Jan'15 by considering only the articles in English and using the default Word2Vec setting: 300 dimensions, 5 words window size, 5 minimum word count (as also used in \cite{mikolov2013distributed}). We also experimented with varied dimension (300, 400, 500, 600, 700) and context size (5, 8, 10) but the results were not significantly improved.
Then, we compare the effectiveness of our approach on ranking the candidate entities with two {\em entity-agnostic} baselines (using the same setting): i) one that considers word embeddings computed from the entire collection of documents (i.e., time-agnostic), and ii) one that considers month-wise word embeddings (i.e., time-aware).\footnote{Note that our approach cannot be compared with \cite{zhang2016probabilistic} because this work addresses the different problem of {\em entity recommendation}, where the input is a {\em free-text query}, not an unambiguous entity URI like in our case.} 

Table \ref{table1} shows the results of normalized Discounted Cumulative Gain (nDCG) \cite{jarvelin2002cumulated} for different top-k lists, without considering time boundary relaxation (i.e., for $w_1=1.0$). Regarding the two entity-agnostic baselines, we notice that the time-aware modeling outperforms the time-agnostic one in all the cases. This improvement is statistically significant for all values of $k$ apart from $k=5$ (paired t-tests, a-level 5\%). 
Regarding the proposed model (time and entity aware), the evaluation results show that it outperforms the baselines in all the cases, while the improvement is statistically significant for all values of $k$. 
This justifies that this model solves the problem of ambiguity that occurs due to the variations of different entity mentions. 

\begin{table}
\vspace{-3mm}
\setlength{\tabcolsep}{5pt}
\renewcommand{\arraystretch}{1.0}
\caption{Evaluation results ($\ddagger$ indicates statistically significant improvement).} \label{table1}
\vspace{-5mm}
\begin{center}
\begin{tabular}{|l|l|l|l|}
\hline
nDCG@k  & Time+Entity Agnostic & Entity Agnostic & Time+Entity Aware \\ 
\hline
k=5 &   0.3210 & 0.3653 & \textbf{0.4999} $\ddagger$ \\ 
\hline
k=10 &  0.3748 & 0.4113 $\ddagger$ & \textbf{0.5402} $\ddagger$\\
\hline
k=20 &  0.4546 & 0.4971 $\ddagger$ & \textbf{0.6115} $\ddagger$\\
\hline
k=30 &  0.5092 & 0.5704 $\ddagger$ & \textbf{0.6562} $\ddagger$\\ 
\hline
\end{tabular}
\end{center}
\vspace{-5mm}
\end{table}

Table \ref{table2} shows the effect of relaxing the time boundaries for 
different $w_1$ values, where $w_2 = w_3$.
In general we see that the effect of relaxing the time boundaries is very small in almost all cases. 
We notice that considering only the query time period $T_q$ provides the best results 
in all cases apart from $k=5$ where $w_1 = 0.9$ slightly performs better.
Moreover, we notice that as the value of $w_1$ decreases (which means increased consideration of $T_{q}^{-1}$ and $T_{q}^{+1}$), the effectiveness of the model gets worse. 
This is an expected result since, for the majority of the query entities in the dataset, the important event related to these entities did not happen very close to the time boundaries. 

An example for which the relaxation of the time boundaries has a positive impact on the ranking is for the entity {\em 2014 FIFA World Cup}. For $w_1 = 0.8$, nDCG@5 increases from 0.45 to 0.51.
Notice that the time period of interest for this entity is July 2014, however the tournament started on June 12, 2014. 
Another example is the query entity {\em Tim Cook} (the CEO of Apple). For $w_1 = 0.8$, nDCG@5 increases from 0.58 to 0.62. Here the query time period is October 2014, however an important event related to Tim Cook happened at the end of the month (he publicly announced that he is gay on October 29). 

We see that the relaxation of time boundaries can positively affect entity relatedness when: i) an important event related to the query entity happened very close to the boundaries of the query time period, and ii) the query entity actually corresponds to an event which spans a long time period. 
Detecting such cases where time boundaries relaxation should be applied is beyond the scope of this paper but an important direction for future work.

\begin{table}
\vspace{-3mm}
\setlength{\tabcolsep}{5pt}
\renewcommand{\arraystretch}{1.0}
\caption{Effect of time boundary relaxation.} 
\label{table2}
\vspace{-5mm}
\begin{center}
\begin{tabular}{|l|l|l|l|l|l| }
\hline
nDCG@k & $w_1=1.0$ &  $w_1=0.9$ &  $w_1=0.8$ &  $w_1=0.7$ &  $w_1=0.6$\\ 
\hline
k=5     &  0.4999   & \textbf{0.5017}   & 0.4990 & 0.4933 & 0.4890 \\ 
\hline
k=10    &  \textbf{0.5402}   & 0.5332   & 0.5358 & 0.5296 & 0.5291 \\
\hline
k=20    &  \textbf{0.6115 }  & 0.6039   & 0.5971 & 0.5932 & 0.5893 \\
\hline
k=30    &  \textbf{0.6562}   & 0.6517   & 0.6451 & 0.6403 & 0.6371 \\ 
\hline
\end{tabular}
\end{center}
\vspace{-5mm}
\end{table}

\section{Conclusion}
\label{sec:conclusion}

We have proposed a flexible model for entity relatedness that considers time-dependent and entity-aware word embeddings by exploiting the corpus of the underlying application. The results of a preliminary evaluation have shown that the proposed approach significantly outperforms similar but time and entity-agnostic models. 

As regards future work, an interesting direction is to extend the proposed method for supporting arbitrary time intervals, which may require joining the results from many models of smaller granularity. However, for supporting very short time periods (e.g., day), this may also require the creation of a large number of Word2Vec models. 
Regarding the relaxation of time boundaries, there is a need of methods that can identify the most reasonable time window to consider given the query entity and time period (by detecting, for example, periods of increased entity popularity \cite{fafalios2017tpdl}). 
Finally, we plan to extensively evaluate the effectiveness of our approach and compare it with state-of-the-art entity relatedness methods using a variety of corpora of different contexts and time periods. 

\section*{Acknowledgements}
The work was partially funded by the European Commission for the ERC Advanced Grant ALEXANDRIA under grant No. 339233.

\bibliographystyle{splncs04}
\bibliography{ESWC_Workshop_EntRel}

\end{document}